\documentclass{elsart}
\usepackage{epsfig}

\begin{document}

\begin{frontmatter}

\title{A Dedicated Computer for Ising-like Spin Glass Models}

\author[Zaragoza,Jarda]{J.~Pech},
\author[Zaragoza]{A.~Taranc\'on} and
\author[Zaragoza]{C.~L.~Ullod} 

\address[Zaragoza]{Departamento de F\'{\i}sica Te\'orica,
	Facultad de Ciencias, \\
	Universidad de Zaragoza, 50009 Zaragoza, Spain}

\address[Jarda]{Institute of Physics, Academy of Sciences, \\
	180 40  Prague, Czech Republic}

\begin{keyword}
Ising model, spin-glass, +/-J, 2d, dedicated machine, programmable logic.
\end{keyword}

\begin{abstract}

We present a parallel machine, based on programmable devices, dedicated to
simulate spin glass models with $Z_2$ variables and short range
interaction. A working prototype is described for two lattices
containing $312 \times 312$ spins each with an update time of 50 {\it ns} 
per spin. The final version of the three dimensional parallel machine is 
discussed with spin update time up to 312 {\it ps}.

\end{abstract}

\end{frontmatter}

DFTUZ/96/19 \hfill{hep-lat/9611014}

\section{Introduction}

Lattice Monte Carlo simulations are an important tool for the physicists 
working in Quantum Field Theory and Statistical Mechanics. These kind of 
simulations require large amounts of computational power and the processing 
can be often parallelized. Therefore, various groups have developed their 
own parallel machines for these simulations
~\cite{APE}~\cite{CHRIST}~\cite{RTN}~\cite{HOOGLAND}.

The appearance in the market of very large  programmable components (PLD)
\cite{PLD} makes it possible to design dedicated machines with low cost 
and high performance. The performance of these machines is increased due 
to the fact that they can run more than a single model: the lattice size 
or the action of the physical model can be easily changed by reprogramming 
the PLD.

At present, spin glass models  \cite{SG}  are an area of Statistical 
Mechanics in progress. They are easily implementable on this kind of 
machines because they  require very simple calculations. These models 
are related to neural networks, spin models, some High $T_c$ 
superconductivity models, etc.

In this paper we describe a PLD-based machine, dedicated to three 
dimensional spin glass models with variables belonging to $Z_2$
and couplings to first and second neighbours.

From the physical point of view, a standard way of studying the model
is by using several independent lattices, called {\it replica}~\cite{SG}.
For this reason we use parallel processing in our machine, running
$n$ independent lattices, in a similar manner to the multi-spin code
used in conventional computers.

An essential tool for the analysis of the results is Finite Size
Scaling ~\cite{FISCHER} which requires the use of different volumes.
Our machine is capable of working with different sizes of lattices by
reprogramming a few components. Also, it is possible to run on a single
large volume considering the {\it n} spins living on the same lattice.
This option is more efficient if the thermalization time is very large.

The algorithm chosen for the simulation is the demon algorithm proposed
by Creutz~\cite{Creutz}. It has been chosen as a starting point because
of its simplicity since it does not require a random number generator for
the update.

At present we have built a bidimensional prototype with first
neighbours couplings. We use this prototype in this paper to explain 
how the machine works and the generalization to the $d=3$ case is 
discussed later.

\section{The Physical Model}

The model we want to simulate is the 3-d Ising-like spin glass model
with first and second neighbour couplings (see ~\cite{SG} for a detailed
description of the model). The action of this model is given by the
expression 

\begin{equation}
S=\sum_{<i,j>} \sigma_i \sigma_j J_{ij} +
  \sum_{<<i,j>>} \sigma_i \sigma_j J_{ij}
\label{action3}
\end{equation}

where the value of the spins $\sigma$ can be only $1$ or $-1$. The same for 
the couplings (multiplied by a constant). While the spins are variables, the 
distribution of couplings is fixed. For a fixed set of $\{J_{ij}\}$ the 
partition function is

\begin{equation}
Z(\beta,\{J_{ij}\})=\sum_{\{\sigma\}} \exp{\beta S(\{J_{ij}\},\{\sigma\}}).
\label{partition}
\end{equation}

To calculate (\ref{partition}) we must sum over $2^V$ possible 
configurations, where $V$ is the volume of the lattice, which is a very large 
number for any computer. The standard way to compute the partition function is
to run an algorithm that selects only a representative set of 
configurations. There are different algorithms to do that. See for 
instance the Sokal chapter in ~\cite{QFC}. For pure spin systems (all 
$J_{ij}$ equal to $1$) some {\it cluster} algorithms are very efficient, 
but for a general {\it spin glass} model, only local algorithms achieve 
good efficiency. Typically we must run the algorithm and generate 
millions of different representative configurations in order to obtain 
accurate results.

The prototype simulates a bidimensional model with nearest neighbour 
couplings only:

\begin{equation}
S=\sum_{<i,j>} \sigma_i \sigma_j J_{ij}.
\label{action}
\end{equation}

The algorithm we use is a microcanonical one. It keeps constant the 
sum of the energy of the lattice and a demon. In order to generate 
the representative set of samples, we start from a spin configuration 
with an action $S$ and a demon energy equal to zero. 

Now we use the algorithm to change the spins to generate new configurations 
(one for every $V$ updates). The update of a spin is as follows: if the flip 
lowers the spin energy, the demon takes that energy and the flip is accepted. 
On the other hand, if the flip grows the spin energy, the change is only made 
if the demon has that energy to give to the spin.

At this level, the $\beta$ value is missing and can be obtained in two ways. 
We can compute the mean energy of the demon over all the sample, 
$\langle E_d \rangle $, and then we obtain the $\beta$ value as 

\begin{equation}
\beta= \frac{1}{4}\log(1+\frac{4}{\langle E_d \rangle}).
\label{Ed1}
\end{equation}

Also if the probability distribution of the demon energy is computed,
$p(E_d)$, the behaviour of this function is given by the expression

\begin{equation}
p(E_d)\propto \exp{(-\beta E_d)}.
\label{Ed2}
\end{equation}

A fit to this function provides us the $\beta$ value.

For simplicity, in case of the prototype machine we use periodic boundary 
conditions in one direction and helicoidal in the other (see 
paragraph ~\ref{first}). The biggest lattice we can simulate has a size 
of 312 by 312 spins. We process 
two independent lattices as an example of how the parallelism can be 
implemented. For spin glass systems these independent configurations 
represent two replicas. By reprogramming some components we can 
simulate either two smaller lattices or a single larger lattice
($624 \times 312$, or similar geometries).

\section{Operation and General Structure of the Machine}

We now describe how the machine performs the simulation.

We assign numbers to the the sites of a lattice from 0 to $V-1$, 
where $V$ is the volume of the lattice. A sequential update is made 
by selecting the spins in the chosen order (see paragraph ~\ref{first}).

To update the spin, its four nearest neighbours have to be supplied 
to the updating engine, too. Then both the local energy $E_\sigma$
and the local energy with the flipped spin $E_{-\sigma}$ are computed.  
The energy balance of the flip $\Delta E= E_{-\sigma} -E_\sigma$ is 
consequently computed to decide whether the flip is accepted:

\begin{itemize}

\item{}If $\Delta E \geq 0$ then the flip of the spin is accepted and 
the demon energy $E_d$ is increased in $\Delta E$.

\item{}If $\Delta E < 0$, but $|\Delta E| \leq E_d$, the flip is 
accepted and the demon energy decreases in $|\Delta E|$.

\item{}Otherwise the flip is rejected and the spin does not change 
its value.

\end{itemize}

These steps are repeated for each spin of the lattice. In order to obtain
an updated spin every clock cycle we have designed a pipeline structure
that performs the latter algorithm step by step.

For all lattices processed in parallel, the spins to be updated 
as well as their neighbours can be stored in memory in such a way 
that each bit of the memory word belongs to a different lattice.
The calculation of the energy balance is independent for each lattice, 
so this part of the hardware must be repeated for any of the lattices 
processed in parallel. 

Now, we give a brief survey of the structure of the spin machine that
performs the previous algorithm. The machine (let us call it SUE, 
for Spin Updating Engine) is connected to a Host Computer (HC).  SUE 
performs the update of the configurations and the measurement of some local 
operators, such as the energies and magnetizations. The rest of measurements 
are  made by HC. Anytime a complete configuration has been updated, HC can 
read the values of the demons from SUE and periodically, after a certain 
number of iterations, SUE is stopped and the configuration is downloaded 
to HC. 

\begin{figure}[t]
\vglue 0.3cm
\epsfig{figure=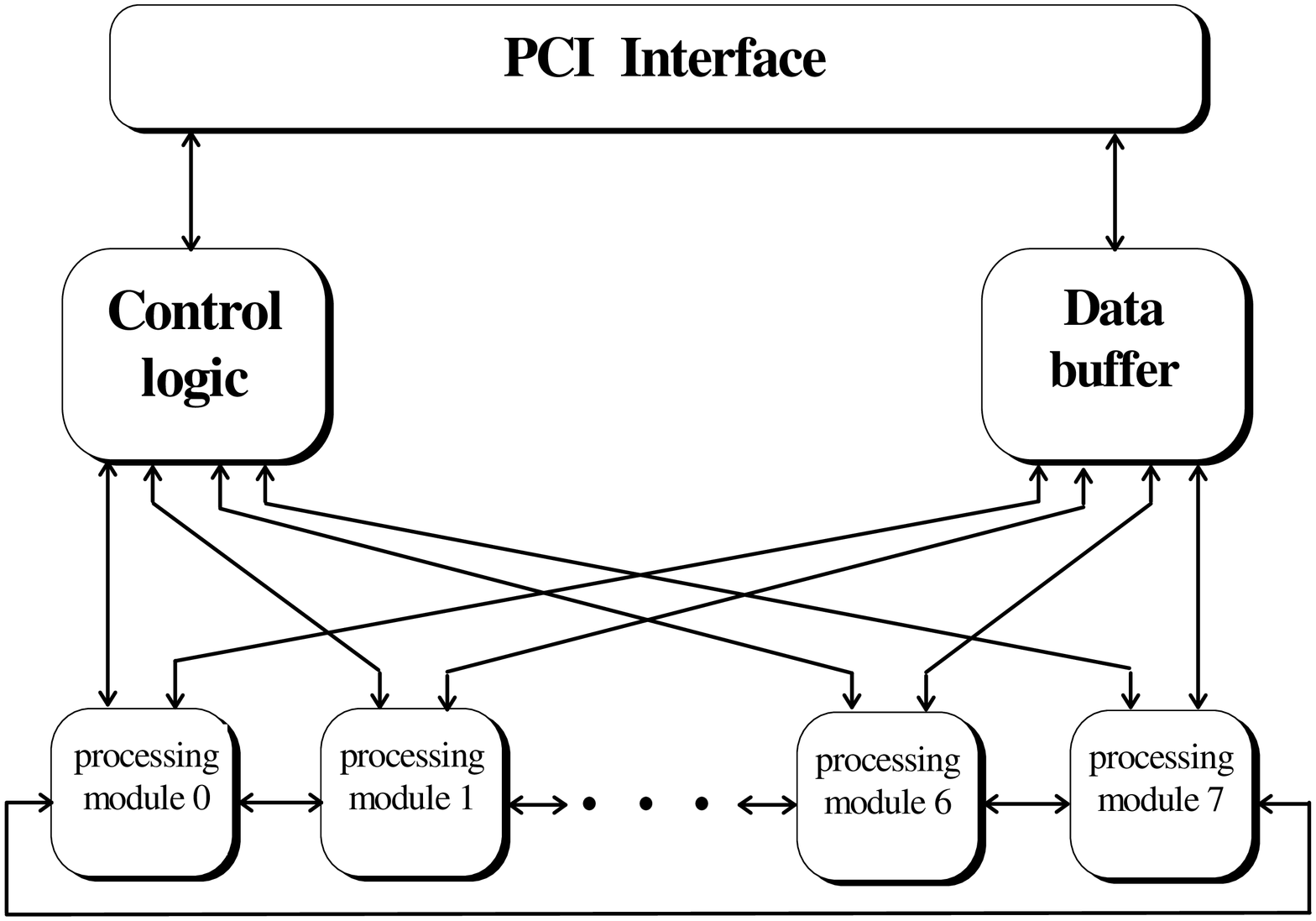,angle=270,width=140mm}
\caption{Block diagram for the $d=3$ machine.}
\label{bl}
\end{figure}

Fig.~\ref{bl} shows a simple diagram of the machine which consists
of a motherboard equipped with $n$ slots for processing modules (the figure
is for $n=8$), PCI interface and control logic. The motherboard provides 
power supply distribution, data interconnection, and allows HC to control 
the processing modules via the PCI interface which is indispensable to 
perform data transfers from/to the modules. Every processing module contains 
the hardware to store and update a set of lattices in parallel. Note that
there are two degrees of parallelism: inside the processing module and 
between the modules.

By processing 8 spins in parallel on 8 modules (64 spins in total) within 
one clock cycle (clock period of 50 MHz), we obtain an update speed of 
312 {\it ps/spin}, a performance more than one order better than 
that of the supercomputers available today.

\section{The $d=2$ Prototype.}

The prototype we have built shows how a processing module works, its
inner architecture and the placement of the algorithm of the simulation
in electronic devices. It also contains the input/output logic which can
be handled from a host equipped with a data acquisition card.

\eject

\begin{figure}[h]
\vglue 0.3cm
\epsfig{figure=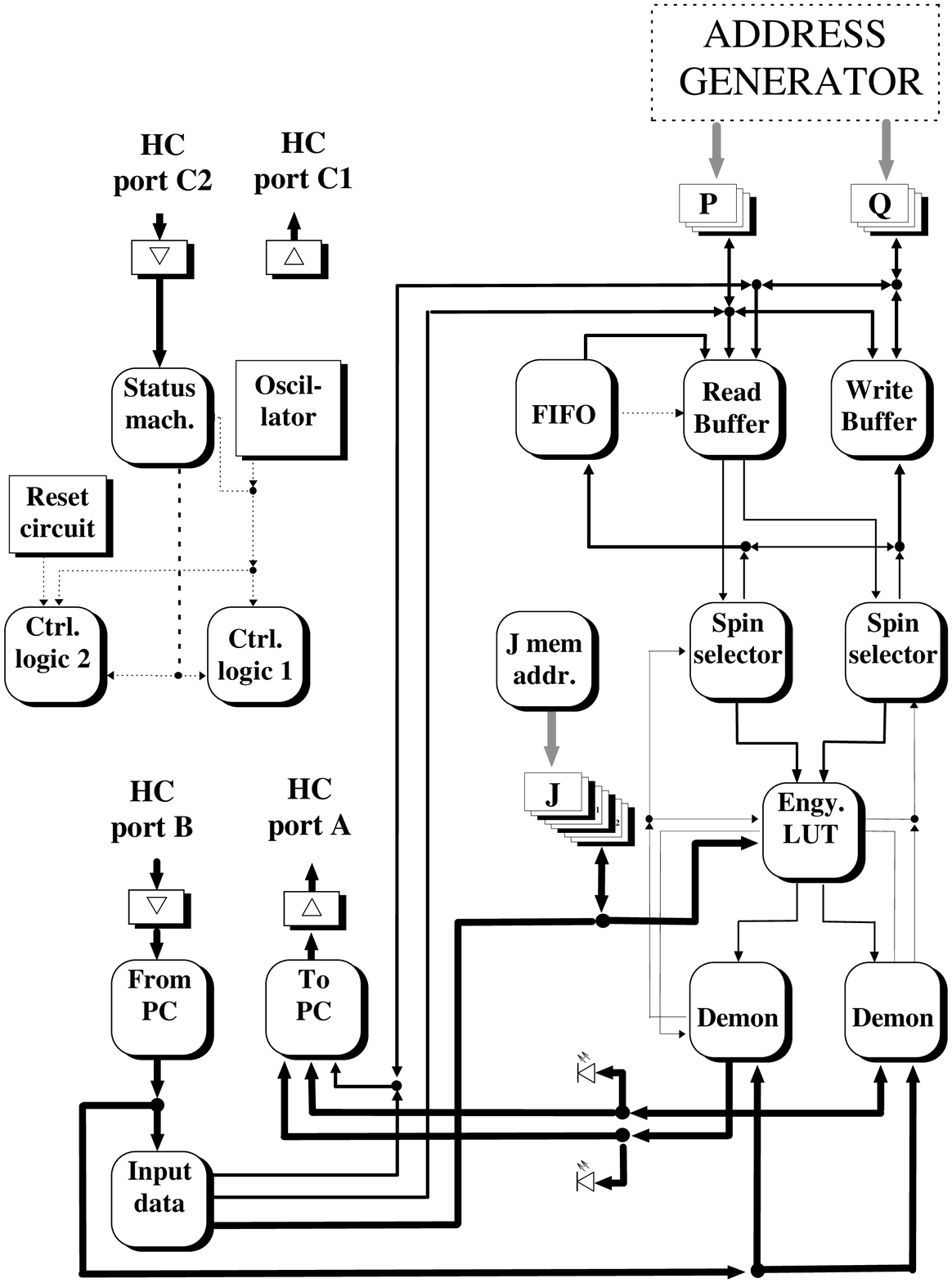,angle=0,width=140mm}
\caption{Block diagram for the $d=2$ prototype.}
\label{layout}
\end{figure}          

\eject

We have used the PLD's Altera EPM7032-10 \cite{PLD} to build the 
logic and the static RAM's (SRAM) KM62256-8. The speed grade of 
the SRAM limits the maximum frequency of operation of the machine to 
10 MHz. SUE performs two updates at every clock cycle, one for each 
lattice, so the theoretical performance is that a spin is updated in 
50 ns. We have designed a pipeline structure (discussed later),  in 
order to obtain an updated spin by cycle. Fig. \ref{layout} shows the 
block diagram of the board: every square is an Altera programmable chip, 
the overlapped rectangular boxes are memory chips and the daughter board 
is an address generator (see paragraph~\ref{third}) that contains seven 
Altera chips.

The logic can be divided into five groups: addressing (daughter board),
spin selection, update, control and I/O logic. 

The {\it addressing logic} prepares addresses for the fetched and stored
(updated) spins. The core of the addressing logic is a set of multiplexed 
counters which provide the address for the spin memory. 

The {\it spin selection logic} contains a subset of the lattice. It 
selects the spin to be updated and its neighbours and sends them to the 
update logic. Each clock cycle it receives an updated spin from the 
update logic.

The {\it update logic} takes the selected spins, the couplings between 
them, and the demon energy from an 8-bit register and carries out the 
update algorithm, sending the new spin to the spin select logic. 

The {\it control logic} is a state machine that handles the set of internal 
signals during the simulation and during the data transfer periods, and 
in addition to the {\it input/output logic} it allows the reading and 
writing of the memory chips and the demon registers, starting and stopping 
the machine.  

The following subsections explain in detail step by step the function of the 
different devices of the prototype. 

\subsection{Spin Storage in Memory}
\label{first} 
We want to store a square lattice of side L, with spin positions labelled 
by $(x,y)$. Let L be a multiple of 3. That is because we  store the whole 
lattice into three memory chips of 32k words.  We use the least significant
bit of the word to store the first lattice and the next one to store 
the second one. Then each address contains  three spins of a lattice, one 
from each memory chip. 

The storage procedure is as follows: every column in the lattice is divided 
into blocks of three spins each (hereafter, the block will be the basic unit 
for labelling the lattice); the first spin of any block is stored in chip 0, 
the second spin in chip 1 and the last spin in chip 2, but the important 
fact is that the generated addresses are the same for all the chips and 
depend only on the block label. We number the blocks vertically in the XY 
plane, beginning from 0 on the top left corner, and then down in the X 
direction. The last block in the first column is the $L/3-1$. We continue 
with the second column and so on. The last block of the lattice has the number
$L^2/3-1$. The individual spins are numbered in the same way, from 0 to
$L^2-1$, and this is the chosen order for carrying out the sequential update.

We present here a nomenclature which will help to clarify the idea. Let us
also tag the spins according to the position they occupy in memory, in the 
way $[chip,address]$. Notice that here we use $[,]$ and for the position 
in the physical lattice we use $(,)$. There is a relationship between 
these two nomenclatures:

\begin{equation}
{(x,y)} = {[x \bmod 3 , y(L/3)+x/3]}.
\label{mapping}
\end{equation}
 
\begin{figure}[t]
\epsfig{figure=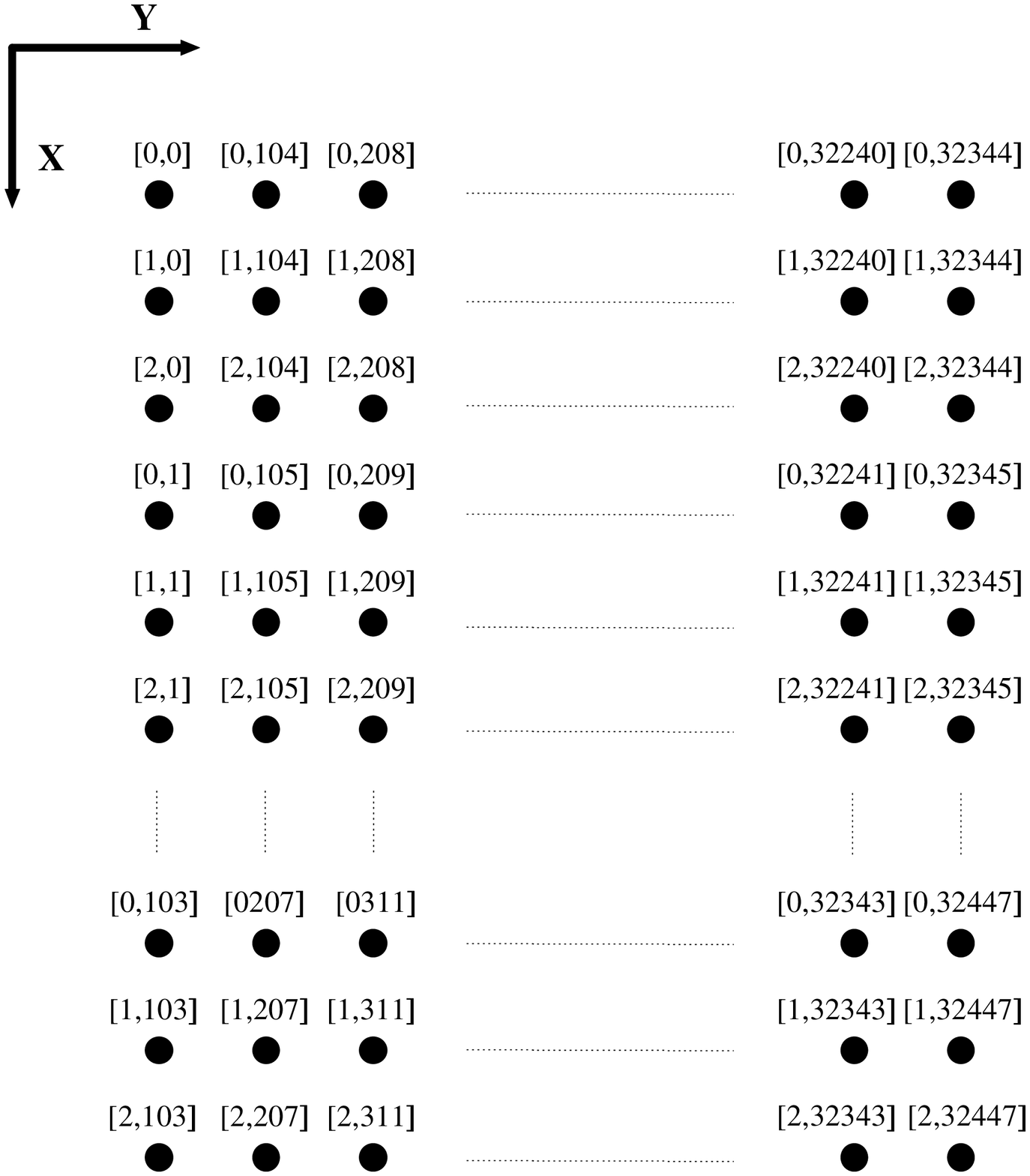,angle=0,width=140mm}
\vglue -2cm
\caption{$L=312$ lattice storage in memory}
\label{ret}
\end{figure}

As said above, the address where a spin is stored is actually the number 
of the block in which this spin can be found. Fig.~\ref{ret} shows the 
distribution of spins in memory for the case of a lattice with
$312 \times 312$ spins. 

In order to write the new spin values during the same read cycle and for 
doing this in a completely automatic manner, the solution is to duplicate
the memory. Then we have two banks of memory (three chips each) which we 
call BankP and BankQ. We use the following update procedure. We want to 
update a column of spins which are read from a bank (i.e. BankP) with the 
neighbour columns. The updated spins are written into the second bank (BankQ).
When the column update is finished, the role of the banks is changed for the 
next column: we read the lattice with the recently updated spins from BankQ 
and the newly updated  spins are stored into BankP. The neighbour columns
are written too (see paragraph~\ref{second}), and due to the fact that an 
updated column is a neighbour column for the next one, the two banks 
finally contain the updated configuration.

The {\it Read Buffer} device (Fig.~\ref{layout}) has access to the spin 
memory data lines. The {\it Write Buffer} chip generates a parity bit which 
is stored together with the updated spins during the write cycles and it is 
checked during the read accesses. The update process takes 8 clock cycles 
from the instant the spin is read until it is updated, so a small 
buffer {\it FIFO} is required in order to bridge over the change of the role 
of the memory banks. This component stores the first updated spins of the 
column being processed.

\subsection{Spin Selection Logic}
\label{second}
The read blocks corresponding to a lattice are written in a sequential way 
in the {\it Spin Selector} device which contains $6$  registers (3-bit wide)
to store  a subset of the lattice. Table~\ref{spinsel} shows these registers:

\begin{table}[t]
\caption{{\it Spin Selector} Registers.}
\label{spinsel}
\centering
\begin{tabular}{ccc}
\small
$[0,A]$      &$[0,B]$     &$[0,C]$   \\  
$[1,A]$      &$[1,B]$     &$[1,C]$   \\  
$[2,A]$      &$[2,B]$     &$[2,C]$   \\  
$[0,D]$      &$[0,E]$     &$[0,F]$   \\  
$[1,D]$      &$[1,E]$     &$[1,F]$   \\  
$[2,D]$      &$[2,E]$     &$[2,F]$   \\  
\normalsize
\end{tabular}
\end{table}

At every moment we have a copy of a certain region of the lattice 
($3\times6$ spins) 
in these registers. Only spins in registers B and E will be updated, moving 
from 0B through 2E. In order to update a spin situated in a position of these 
registers, its neighbours must be correctly placed in the rest of the 
registers. This component contains a state machine that performs repeatedly 
6-step loops with the following structure:

\begin{itemize}

\item
Step 0: 
\begin{itemize}
\item
Register A is sent to the Write Buffer and replaced by the following block.
\item
Spin 0E is sent to the update logic.
\item
Spin 1B updated is received.
\end{itemize}
\item
Step 1: 
\begin{itemize}
\item
Register B (the updated spins) is sent to the Write Buffer and replaced 
by the following block.
\item
Spin 1E is sent to the update logic.
\item
Spin 2B updated is received and sent to the Write Buffer with the other two
spins (already updated) of its block.
\end{itemize}
\item
Step 2: 
\begin{itemize}
\item
Register C is sent to the Write Buffer and replaced by the following block.
\item
Spin 2E is sent to the update logic.
\item
Spin 0E updated is received.
\end{itemize}
\item
Step 3:
\begin{itemize}
\item
Register D is sent to the Write Buffer and replaced by the following block.
\item
Spin 0A is sent to the update logic.
\item
Spin 1E updated is received.
\end{itemize}
\item
Step 4: 
\begin{itemize}
\item
Register E (the updated spins) is sent to the Write Buffer and replaced 
by the following block.
\item
Spin 1A is sent to the update logic.
\item
Spin 2E updated is received and sent to the Write Buffer with the other two
spins (already updated) of its block.
\end{itemize}
\item
Step 5: 
\begin{itemize}
\item
Register F is sent to the Write Buffer and replaced by the following block.
\item
Spin 2A is sent to the update logic.
\item
Spin 0A updated is received.
\end{itemize}
\end{itemize}

The order in which blocks are read from memory and loaded into the 
corresponding Altera register is given in the table~\ref{order}, from
left to right and top to bottom.

\begin{table}[h]
\caption{Block reading order.}
\label{order}
\centering
\begin{tabular}{ccc}
C0       &C1        &C2        \\
\hline
$0$      &$104$     &$208$   \\  
$1$      &$105$     &$209$   \\  
$2$      &$106$     &$210$   \\  
$3$      &$107$     &$211$   \\  
$4$      &$108$     &$212$   \\  
$5$      &$109$     &$213$   \\  
$6$      &$110$     &$214$   \\  
$\dots$  &$\dots$   &$\dots$   \\
$32446$  &$102$     &$206$   \\  
$32447$  &$103$     &$207$   \\  
$0$      &$104$     &$208$   \\  
$1$      &$105$     &$209$   \\  
\end{tabular}
\end{table}

As shown in table~\ref{order}, the three columns contain memory locations
with the spin to be updated (column C1)  and its neighbour spins. All three 
columns (one line) has to be fetched before the update procedure of the spins
can be started. In this way, all memory locations are read and updated. The 
blocks in column C0 feed the A and D registers, in column C1 the B and E 
registers, and in column C2 the C and F registers. The spins to be updated 
are always loaded into either the B or E registers. Remark that the previously
updated column feeds the A and D registers.

\subsection{Addressing Logic.}
\label{third}

As can be seen from table~\ref{order}, it is very simple to generate the
sequence of addresses to select the blocks which are read from memory, and
the order in which the blocks are written into memory. The order is the same,
but shifted a few cycles. For this reason, it is adequate to have two sets of
three counters (C0, C1, C2): one for reading and the other for writing. They 
correspond to the columns in table ~\ref{order}, so C0 begins to count from 0,
C1 begins from 104 and C2 begins from 208, and they pass through all the 
values from 0 to $L^3/3-1$. 

These counters and their multiplexes are programmed in a 
{\it Daughter Board} which is a plug-in module of SUE.

\subsection{Update Logic.}

The update logic takes the spin to be updated and their neighbours in order
to calculate the amount of energy that the flip requires. One of the neighbour
spins arrives directly from the update logic. The four couplings that link 
the spins are read from a memory bank (J). This memory is addressed by the 
{\it J Mem. Addr.} device. This bank is $3\times32k$ deep by $9$ bits wide. As
only four bits are used for one lattice, two sets of couplings for two 
different lattices can be stored in a single memory word. We have also
incorporated a parity bit which is checked by the {\it Energy Look Up Table 
(LUT)} device. The update procedure is carried out in two clock cycles. 
During the first cycle the spins and their relative couplings are fed into the 
{\it Look Up Table} whose output is the energy balance of the update. Along 
the second cycle this value is added to the demon energy in the device called 
{\it Demon} and the sign of the sum is checked. If this sign is greater than 
or equal to zero, the flip is accepted and the demon changes its value. If 
the sign is lower than zero, the spin and the demon do not change. The updated
spin is fed back to both the LUT and spin selection logic.

\subsection{Status Machine.}

The way in which the SUE works is programmed as a big status machine in the
{\it Status Machine} component. This chip receives the instructions released
by HC and together with the {\it Control Logic} chips it generates a set of 
signals required to control the memories, buses, etc.

\subsection{Input/Output Logic.}

HC is connected to SUE via a data acquisition board based on 8255A-like
controller with interrupt request (IRQ) capabilities. We use two 8-bit 
ports (A,B) and two 4-bit ports (C1,C2). A and C1 ports are output for SUE 
while B and C2 are input ports. The 8-bit ports are used for data 
transmissions, and C2 is used to give instructions to SUE. The signals in C1 
are the IRQ, the reset signal and the two parity errors checked. 
{\it From PC, To PC} and {\it Input Data} components allow HC to access to 
the different memory banks and registers of SUE.

The following set of instructions is available through the port C2:
\begin{itemize}
\item Reset.
\item Read/write demon energy for the first lattice.
\item Read/write demon energy for the second lattice.
\item Read/write a spin block (the two lattices at the same time).
\item Write a coupling word.
\item Start simulation.
\end{itemize}

The normal operation of SUE is as follows:
\begin{itemize}
\item Store couplings in J-memory.
\item Store spins in spin memory.
\item Write the initial demons' energies.
\item Start simulation.
\end{itemize}

When the start instruction is executed, SUE reads a number $n$ from the data
acquisition board and it begins to generate $2^n$ updates of the initial
configuration. Every time an update of the configuration is completed, SUE
generates an IRQ to HC and the demon registers can be read by HC. When the 
$2^n$ updates are performed, the SUE stops and HC can download the spin 
configurations. In order to restart the simulation it is not necessary 
to rewrite the configurations, but to execute again a start instruction. Data 
transmission speed through the ISA DAQ card is 2 kBytes/s, so storing a 
configuration of spins takes $0.15 sec$.

\section{Design Considerations and Final Product}

As it was formerly said, for the prototype version we have used
the PLD's Altera EPM-7032-10. The logic has been designed with registered 
logic and short propagation delay times, so the highest frequency of 
operation is fixed by the memory access time to 10MHz. In these conditions, 
we obtain two updated spins in 100 ns. Nevertheless, the reliability of 
the double side printed circuit board allows us to work at 5MHz only. 
Consequently, the real update speed of this machine is 1 updated spin 
in 100 ns.

We can compare SUE performance versus a general purpose computer: We have 
written an optimized program that runs the same model, simulating 8 lattices 
at the same time. With this degree of parallelism, a 120 MHz Pentium PC 
takes 1000 ns in order to update 1 spin.

\section{Physical Results}

In order to check the SUE performance and their reliability we have
run a simulation on the critical point of the Ising model. The exact
(analytical) solution of this model is known and so obtaining
the correct value is a very good test for the SUE's global functionality.
We start with a configuration as close as possible to the critical energy 
$S/V =\sqrt{2}/2$. The obtained $\beta$ value from the simulation using 
(\ref{Ed1}) and (\ref{Ed2}) has to be $\beta_c=\frac{1}{2}\log(1+\sqrt{2})$.

\begin{figure}[b]
\vglue 0.3cm
\epsfig{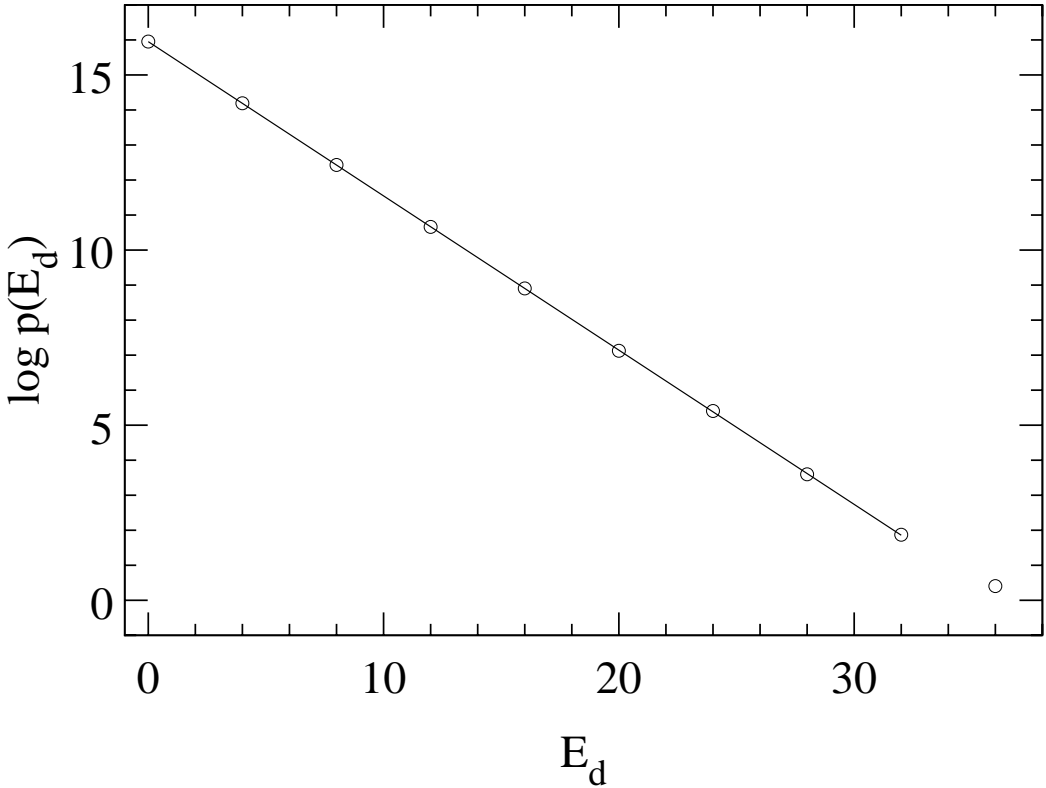}
\caption{Linear fit for the demon energy distribution.}
\label{frec}
\end{figure}          

Due to the fact that the first configuration is very far from the equilibrium
(it is not in the representative sample), we must thermalize it first.
We do that by running $10^6$ iterations. Then we run $2 \times 10^6$
iterations to measure. After every iteration (update of $V$ spins) SUE outputs
the demon energy to HC. Every 1024 iterations the full configuration is 
downloaded to HC and checked in order to control that the global energy has 
not changed.

From (\ref{Ed1}) we obtain $\beta =0.44057(7)$ to be compared with
the exact value $\beta_{exact}=0.44068\dots$. The small difference
is due to the Finite Size Effects because the exact $\beta$ is only
obtained in the limit $V\to\infty$. In this case, the shift in
$\beta$ \cite{FERDI} is

\begin{equation}
\beta_\infty - \beta_L \approx {-0.16 \over L} = 5 \times 10^{-4} 
\label{dbeta}
\end{equation}

which makes results compatible.

To obtain $\beta$ from (\ref{Ed2}) we plot the probability distribution of 
the demon energy in fig (\ref{frec}). With a linear fit to the first $9$ 
points we obtain $\beta=0.4404(15)$, which is correct with a very larger
error than in the previous case. The straight line in the figure is the 
final fit.

\section{The $d=3$ Generalization} 

The three-dimensional machine should be based on the same philosophy
as the currently working prototype for $d=2$. The {\it Spin Selector} has 
to be extended in order to store a greater region of the lattice. A simple 
extension is described here.

For the three-dimensional lattice we have to use blocks of 9 spins each,
keeping the same numbering system of blocks as before, plane by plane. The 
first plane has blocks from 0 to $L^2/9 -1$. The next plane has its blocks 
numbered from $L^2/9$ to $2L^2/9-1$. The last plane begins in the block
$(L-1)L^2/9$ and ends in the $L^3/9-1$ block.

\begin{figure}[h]
\vglue -2cm
\epsfig{figure=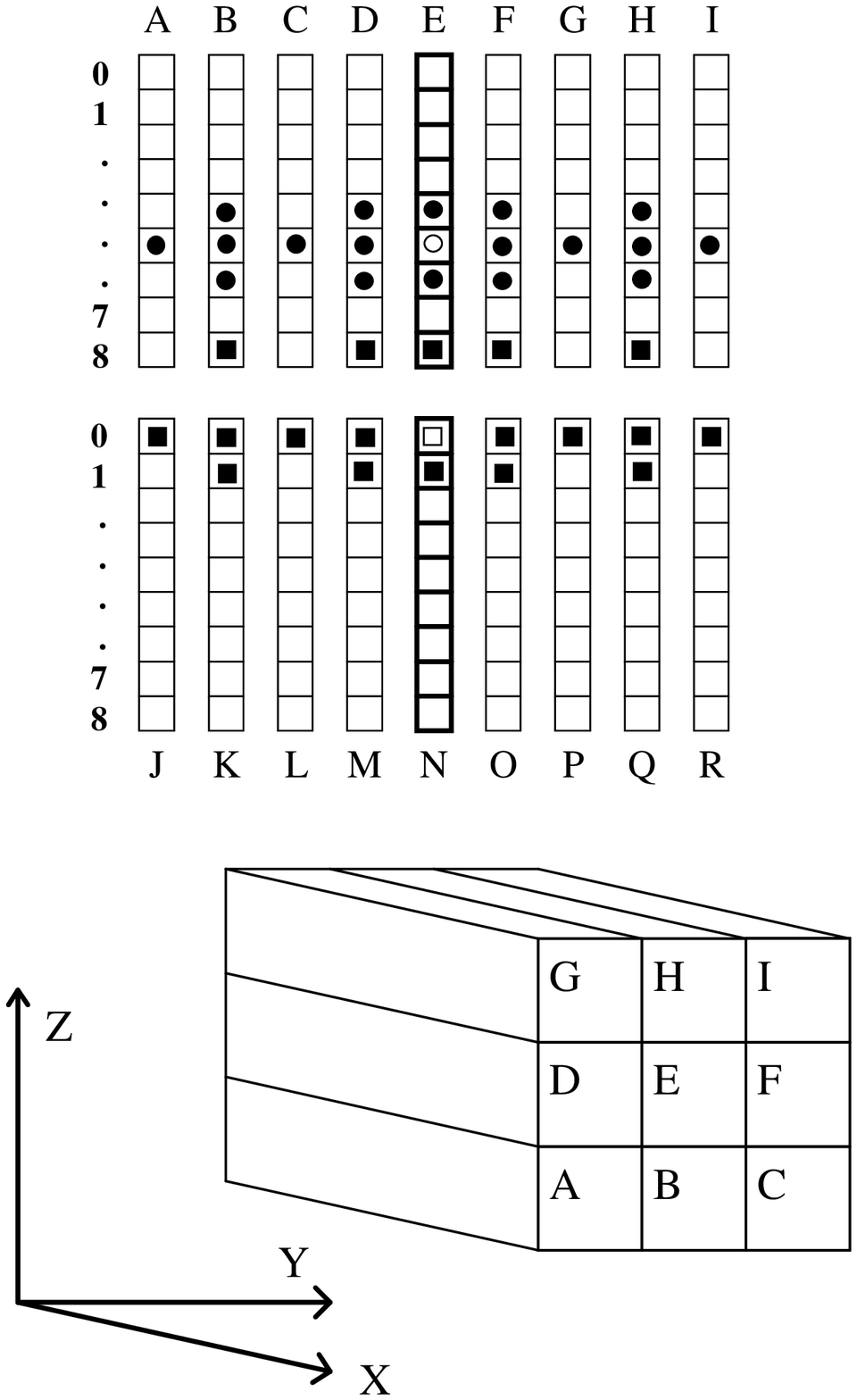,angle=0,width=140mm}
\vglue -2cm
\caption{Spin selector for the $d=3$ case.}
\label{3dsel}
\end{figure}

Fig.~\ref{3dsel} shows the extension of the {\it Spin Selector} logic.
We use 18 registers, 8 bits each: the central registers (E,N) contain 
the spins to be updated and the rest of the registers store the neighbours:
The D (M), E (N) and F (O) registers contain the neighbours in the same $XY$ 
plane; The A (J), B (K) and  C (L) registers store the neighbours in the
plane below, and the G (P),H (Q) and I (R) registers store the neighbours
corresponding to the $XY$ plane above. The neighbours of two spins have been 
depicted in the figure, besides the localization of the 9-tuples in the 
lattices. The spins to be updated are selected along the $X$ axis. Remark
the boundary conditions between the two sets of 9-tuples.

The addressing logic has to provide addresses in such a way that the 
registers are fed in alphabetical order. For instance, when the spin $E_0$ 
is sent to the update logic, the block $J$ is read from memory. This component
will perform a loop of 18 different states, like the 6-state loop for the 
bidimensional case.

The addressing logic keeps the structure of multiplexed counters. The 
boundary conditions are periodical in one direction and helicoidal
in the other two directions.

\begin{figure}[t]
\vglue 0.3cm
\epsfig{figure=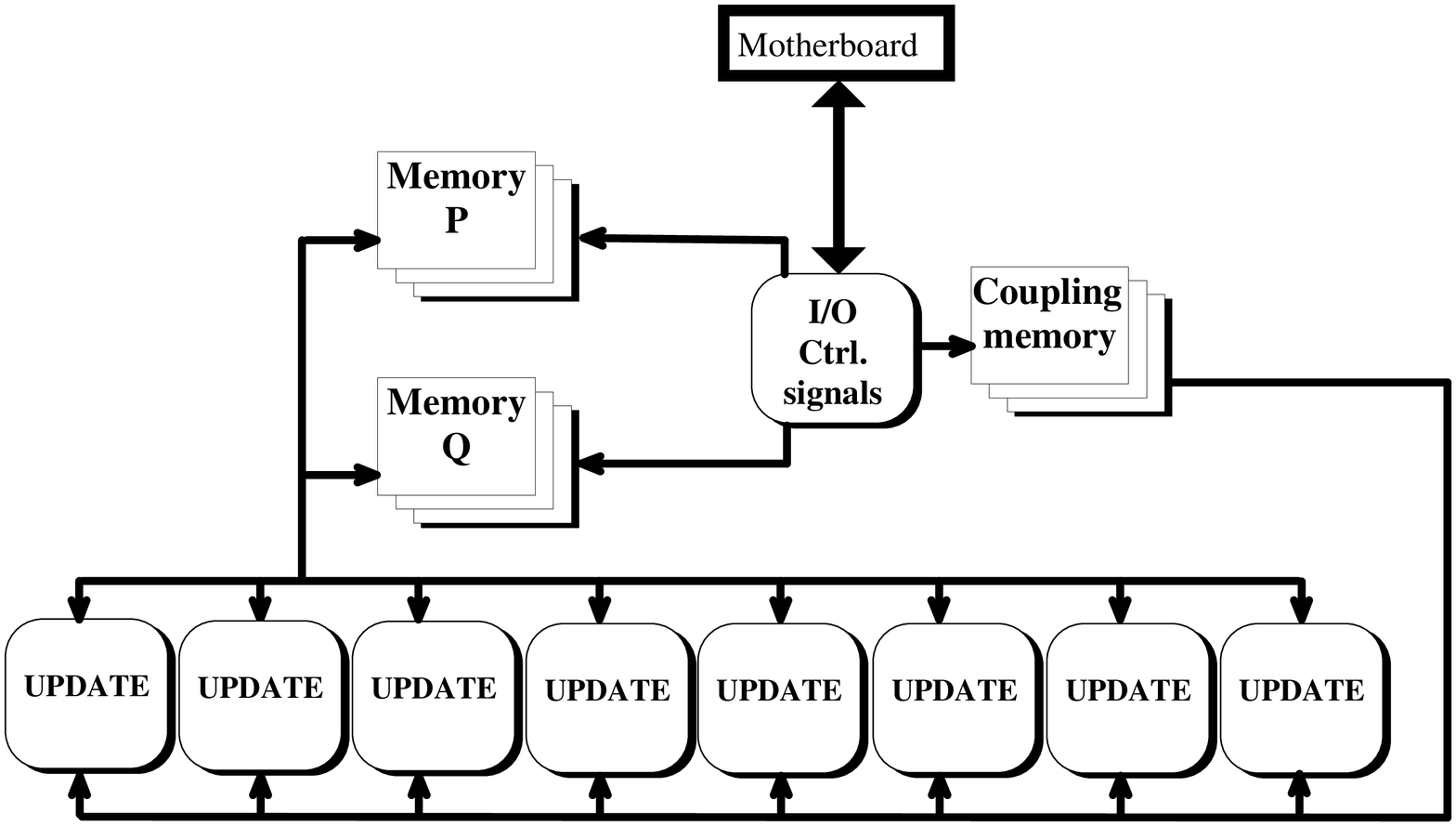,angle=270,width=140mm}
\vglue -1cm
\caption{Block diagram for a $d=3$ processing module.}
\label{3d}
\end{figure}          

For the construction of this board the high capacity Altera PLDs' (CPLD) 
and fast Static RAMs (with $20 ns$ access time) has to be used. In this way, 
each of the two memory banks P and Q will consist of from nine $128K \times 8$
SRAM components. The 8-bit word is used in order to run 8 independent
lattices and 9 memory components to build the 9-bit blocks. The largest 
symmetric lattice that can be stored with $L$ multiple of 9 is $99 \times
99 \times 99$ spins. The I/O and control logic will be placed in a single
PLD and the {\it Spin Selector} and {\it Update Logic} into another PLD.
Modules bearing the latter PLDs allow us further segmentation of the lattices
and to speed up the spin update time (see fig.~\ref{3d}). 

As the memory access time determines the maximum clock frequency
(50 MHz), this board can make an update in ${{20 \,\mbox{\it ns}\over {8}}
= {2.5 \,\mbox{\it ns}}} $. A motherboard with 8 plug-in modules
can reach an update time of 312 {\it ps} per spin.

We also want to include into the latter PLD's a random number generator 
in order to run a canonical simulation. The random number generation is a
time consuming task in a general purpose machine, making slow the
simulation. In a dedicated machine this generation can be carried out
without time loss, raising its performance with respect to conventional
computers.  

We wish to thank J.~Carmona, D.~I\~niguez, J.~J.~Ruiz-Lorenzo, 
G.~Parisi and E.~Marinari for useful discussions. Partially supported 
by CICyT AEN93-0604-C01 and AEN94-0218. CLU is a DGA Fellow.

\end{document}